\documentclass[proof]{WileyASNA-v1}

\articletype{Article Type}%

\received{26 April 2018}
\revised{6 June 2018}
\accepted{21 June 2018}

\begin{document}

\title{Studying the variability of the X-ray spectral parameters of high-redshift GRBs' afterglows
}

\author[1,2,3]{Istvan I. Racz*}

\author[1]{Agnes J. Hortobagyi**}

\authormark{RACZ \textsc{et al}}

\address[1]{\orgdiv{Department of Physics of Complex Systems}, \orgname{E\"otv\"os Lor\'and University}, \orgaddress{\state{Budapest}, \country{Hungary}}}

\address[2]{\orgdiv{Department of Natural Science}, \orgname{National Univesity of Public Services}, \orgaddress{\state{Budapest}, \country{Hungary}}}

\address[3]{\orgdiv{Konkoly Observatory, Research Centre for Astronomy and Earth Sciences}, \orgname{Hungarian Academy of Sciences}, \orgaddress{\state{Budapest}, \country{Hungary}}}

\corres{*\email{racz@complex.elte.hu}\\ **\email{ajhortobagyi@caesar.elte.hu}}

\presentaddress{5-th floor, office: 5.80, 1/A P\'azm\'any P\'eter s\'et\'any, 1117 Budapest, Hungary}

\abstract{
The Swift satellite has observed more than a thousand GRBs with X-ray data. Almost a third of them have redshift measurement, too. Here we start to investigate the X-ray spectral fitting of the data considering the low energy part where the N(H) absorption happens. Based on the available more accurate input data we examined the robustness of previous fittings and tested how sensitive the changes of the starting parameters are. We studied the change of the intrinsic hydrogen column density during the outburst for a few events. No significant variability of N(H) column density was identified.}


\jnlcitation{\cname{%
\author{Istvan I. Racz}, 
and \author{Agnes J. Hortobagyi}
} (\cyear{2018}), 
\ctitle{Studying the variability of the X-ray spectral parameters of high-redshift GRBs' afterglows}, \cjournal{AN}, \cvol{2018;XX:X--X}.}


\maketitle


\section{Introduction}

Gamma-ray bursts (GRBs) are the brightest explosions that have been observed in the distant Universe (see the review paper of \citet{kumar}). GRBs can be observed at all wavelengths, from gamma-rays to radio \citep{amati,meszaros}.
The duration of GRBs can be characterized by the $T_{90}$ -- or  less commonly by $T_{50}$ -- which is the time taken to accumulate 90\% (50\%) of the total observed photons. 	
Several thousand GRBs have been detected so far, but only about 500 have measured distance/redshift \citep{zsolt2003_redshift2,zsolt2003_redshift,zsolt2006_redshift}. 

There are 2 major groups of GRBs based on their durations: short and long. However, there exsists evidence for the existence of a third group: intermediate. \citep{hoi1998,hoi2002,hoi2006,hoi2008,pista_2_inter,ugarte,pista1_inter}. The mean durations of these groups depend on the observing detector, e.g. for the Swift BAT they are $\rm{T_{90}\approx0.3\,s}$, $\rm{T_{90}\approx8.5\,s}$, and $\rm{T_{90}\approx40\,s}$, respectively \citep{hoi2010}. 
According to the widely accepted theoretical assumptions there are two main central engines producing GRBs: compact cosmic objects merging and the collapse of high-mass stars. The merging of neutron stars or black holes can cause the short GRBs, while the collapsars are responsible for the long GRBs. The recently observed gravitational wave event GRB170817A/GW170817 had an electromagnetic counterpart in the form of a GRB and originated from merging two compact objects of 1.36 and 2.26$M_\odot$. \citet{hoi2018} classified this event as an intermediate-duration object, and pointed out the possibility that the intermediate GRBs originate from neutron stars merging.  The long GRBs are believed to be indicators of star formation and, as such, indicate the large scale structures of the Universe \citep{zsolt1998_pca,balazs1999,zsolt2000_anizotrop,zsolt2000_anizotrop2,pista_3_anizotrop,zsolt2010_anizotrop}. Recently, the Hercules-Corona Borealis Great Wall was found by \citet{hoi2014,hoi2015} and the Giant GRB Ring was found by \citet{ring,ring2}. 

Both central engine models produce collimated high-energy jets responsible for the observed radiation in the gamma-range.  The jet produces two types of emission: the prompt emission and the afterglow emission from its collision with the circumstellar matter \citep{meszaros_2006,kumar}.  The prompt emission basically refers to the operation of the central engine, while the afterglow provides information on the local, intergalactic and galactic medium \citep{starling}. 

Any matter between the detector and the GRB source attenuates the X-rays emitted by the source and modulates the intrinsic X-ray spectrum.  There are three kind of interfering matter: the galactic foreground, the intergalactic matter and the interstellar matter surrounding the source. 
The magnitude of the effect depends on the column density of a given absorber, e.g. as the external shock wave, responsible for the afterglow, is proceeding in the interstellar matter the X-ray absorption is changing.

One may expect a change in the observed column density modulating the intrinsic X-ray spectrum as the jet transmits through the surrounding interstellar matter.  Such effect was observed by \citet{amati} and the aim of our paper is to look for this effect in some bright Swift GRBs. 

The paper is organized in four sections.  In Section~\ref{obs}., we give an overview on X-ray observation for the Swift GRBs. The X-ray spectral fitting tools are shown in Section~\ref{fit}. Then we present the intrinsic N(H) column density evolution in Section~\ref{evolution}. Finally, in Section~\ref{sum}. we summarize our results.



\section{X-ray observation}
\label{obs}

The GRBs used in our calculations were detected by the \textit{Neil Gehrels Swift Observatory} (formally known as 'Swift' space telescope) \citep{bat,xrt,uvot}.  The Swift telescope has 3 different instruments: the Burst Alert Telescope (BAT), the X-ray Telescope (XRT), and the UV/Optical Telescope (UVOT).  The three instruments detect the following energy ranges: 15-150keV for the BAT, 0.3-10keV for the XRT, and 170-600nm for the UVOT. The \textit{Swift BAT}, \textit{XRT} and \textit{UVOT} observe the outbursts in gamma, X-ray, and ultraviolet/optical respectively. The Swift detected ${\ge}\,1350$ GRBs until mid-April 2018.  From these 1350 GRBs the XRT has found more than 970 X-ray afterglow emissions. 

The radiation leaving the GRB travels through the intergalactic and galactic foreground, which significantly affects the spectrum we observe \citep{evans2007}. The gamma-rays usually don't interact noticeably with the interstellar medium but the X-ray and shorter wavelength photons do \citep{wilms,evans2009,kumar}. 

The galactic foreground is not as homogeneous as we have previously thought \citep{toth1,racz}. It seems to be heavily structured and clumpy. The estimation of the foreground base on the radio surveys of atomic hydrogen having an angular resolution of magnitude of degrees. The GRBs are point-like sources, therefore the fine structure of the foreground is very important.



\section{X-ray spectral fitting}
\label{fit}

To determine the X-ray spectrum we used XSpec, which is a widely used X-Ray Spectral Fitting Software Package \citep{xpsec1, xpsec2, xpsec3}. One can set many parameters like cosmological or solar abundances, and use some models constructed from individual components.

The GRBs' X-ray light curves can be well approximated by a sectioned power law function (in time) with breakpoints separating the different phases. The \textit{Swift} data pipeline produces the phases and corresponding X-ray spectra automatically. In this study, the publicly available XRT spectra is used. The UK Swift Science Data Centre (UKSSDC) provides a semi-automatic algorithm for the Swift XRT spectral analysis \citet{evans2007,evans2009}. This study used that method with the standard initial settings and parameters: flat Universe, standard photoionization cross-sections by \citet{vern}, and low metallicity stellar abundances by \citet{wilms}. We used multiplicative models employing the Tuebingen-Boulder ISM absorption models (twice for the foreground and the intrinsic), the simple photon power law, and the convolution model to calculate the flux.

\begin{figure}
\includegraphics[width=1\linewidth]{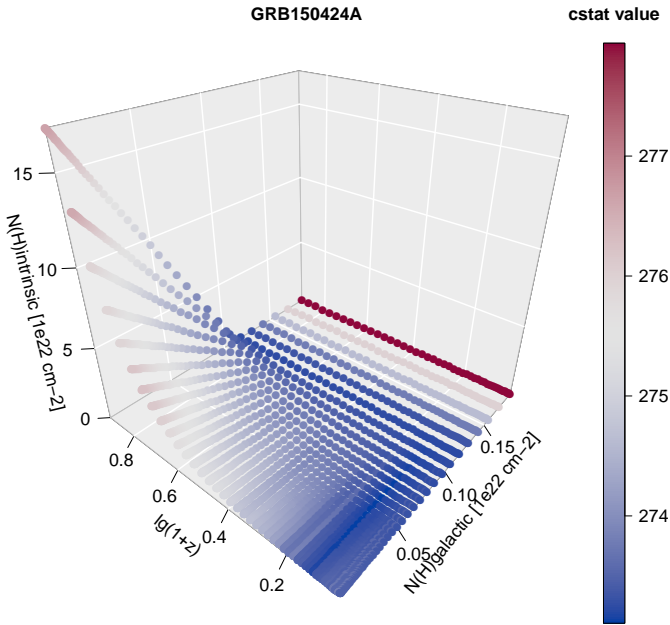}
\caption{The dependence of the intrinsic column density on the N(H) galactic foreground density and on the redshift. No intergalactic N(H) was assumed. The figure shows that there is a linear connection between galactic and intrinsic N(H) densities, while they have a quadratic relationship with the lg(1+z).\label{3d}}
\end{figure}

The X-ray spectrum observed by the satellite is modified by the attenuation of N(H) column density in the line of sight; therefore we examined the dependence of intrinsic column density on each starting parameter and how sensitive the fitting is to these changes. The background and the data contains Poissonic noise, hence we applied the C-statistic process. 
To investigate the impact of the foreground galactic N(H) and the redshift on the intrinsic N(H), we calculated those column densities for the GRB150424A. On Fig.~\ref{3d}. the dependence of the intrinsic column density on the N(H) galactic foreground density and on the redshift are shown.  Results indicate a linear connection between galactic and intrinsic N(H) densities, while a quadratic relationship is apparent with the redshift.

The quadratic relationship can be explained by a simple model, where both the attenuating material and the intrinsic spectral energy distribution shapes the spectra \citep{arnaud}. For an optically thin N(H) column the observed spectral intensity, M(E) will be given by:

\begin{equation}
M(E)=exp[-N(H)\cdot \sigma(E)] \label{ex}
\end{equation}

\noindent where N(H) and $\sigma(E)$ are the hydrogen column density and the photo-electric cross-section, respectively. The $\sigma(E)$ without the "0" index refers to the frame moving with a speed of the observer, relative to the rest frame.

\begin{equation}
\sigma^0(E_0)=\sigma(E_0/(1+z))
\end{equation}





The distances of most GRBs are not determined, the Eq.~(\ref{ex}) shows that in the frame moving with the observers the observed spectral intensity will be: 

\begin{equation}
M(E)=exp[-N(H)^* \cdot \sigma(E)]
\end{equation}

\noindent where

\begin{equation}
N(H)^*=N(H)\frac{\sigma[E(1+z)]}{Q(E)}
\label{exh}
\end{equation}

\noindent One can see that the $N(H)^*$ also depends on $E$ besides fixed column density, the quotient at the right side of Eq.~(\ref{exh}) depends only on $(1+z)^a$. The XSpec software uses the Born approximation \citep{born}, where the $\alpha$ value equals 2.

\section{Evolution of $N(H)_{intrinsic}$}
\label{evolution}

To investigate the $N(H)_{intrinsic}$ changes during the GRBs, we randomly selected multiple GRB samples with unknown redshifts and three breakpoints in the Swift X-ray light curve. There are 4+1 spectra for the 3 breakpoints, because of the two modes of XRT: Window Timing and Photon Counting.

Table~\ref{evo}. shows the selected GRBs and the time range of each phase. Results from GRB160815A and GRB161202A during the first phase demonstrate that Swift was still settling at the time those readings were taken -- therefore these were omitted from the analysis.

\begin{table*}
\centering
\caption{The breakpoints of the 6 selected GRBs. These GRBs have 3 breakpoints in their light-curves and both WT and PC mode spectra, so there are 5 time phases. The columns show the starting and ending times (in seconds) of each phase relative to the trigger. The bold faced numbers show the omitted spectra, where the Swift was still in settling.\label{evo}}
\begin{tabular}{c||r|r||r|r||r|r||r|r||r|r}
\textbf{ID of GRB} & \multicolumn{2}{c||}{\textbf{Phase1}} & \multicolumn{2}{c||}{\textbf{Phase2}} & \multicolumn{2}{c||}{\textbf{Phase3}} & \multicolumn{2}{c||}{\textbf{Phase4}} & \multicolumn{2}{c}{\textbf{Phase5}}\\ \hline
GRB150424A& 94&152 & 152&360 & 360&681& 682&56566 & 114114&1442306\\
GRB150925A& 146&148 & 148&344 & 345&394 & 395&1686 & 5027&93155\\
GRB160412A& 85&108 & 108&248 & 248&378 & 379&12288 & 18684&1236027\\
GRB160815A& \textbf{95}&\textbf{97} & 97&122 & 122&177 & 177&27989 & 27990&189870\\
GRB161202A& \textbf{64}&\textbf{79} & 80&125 & 125&236 & 237&6163 & 9401&160583\\
GRB170330A& 104&200 & 200&368 & 368&800 & 801&11501& 35757&253302\\
\end{tabular}
\end{table*}

One can expect the intrinsic column density to show a decrease in time because the X-ray radiation originates from the interaction between the jet and dropped environment based on previous analysis, e.g. \cite{kumar}.  In this model the jet moves through the medium with relativistic speed, the released X-ray radiation is absorbing less and less hydrogen.

\begin{figure}
\includegraphics[width=1\linewidth]{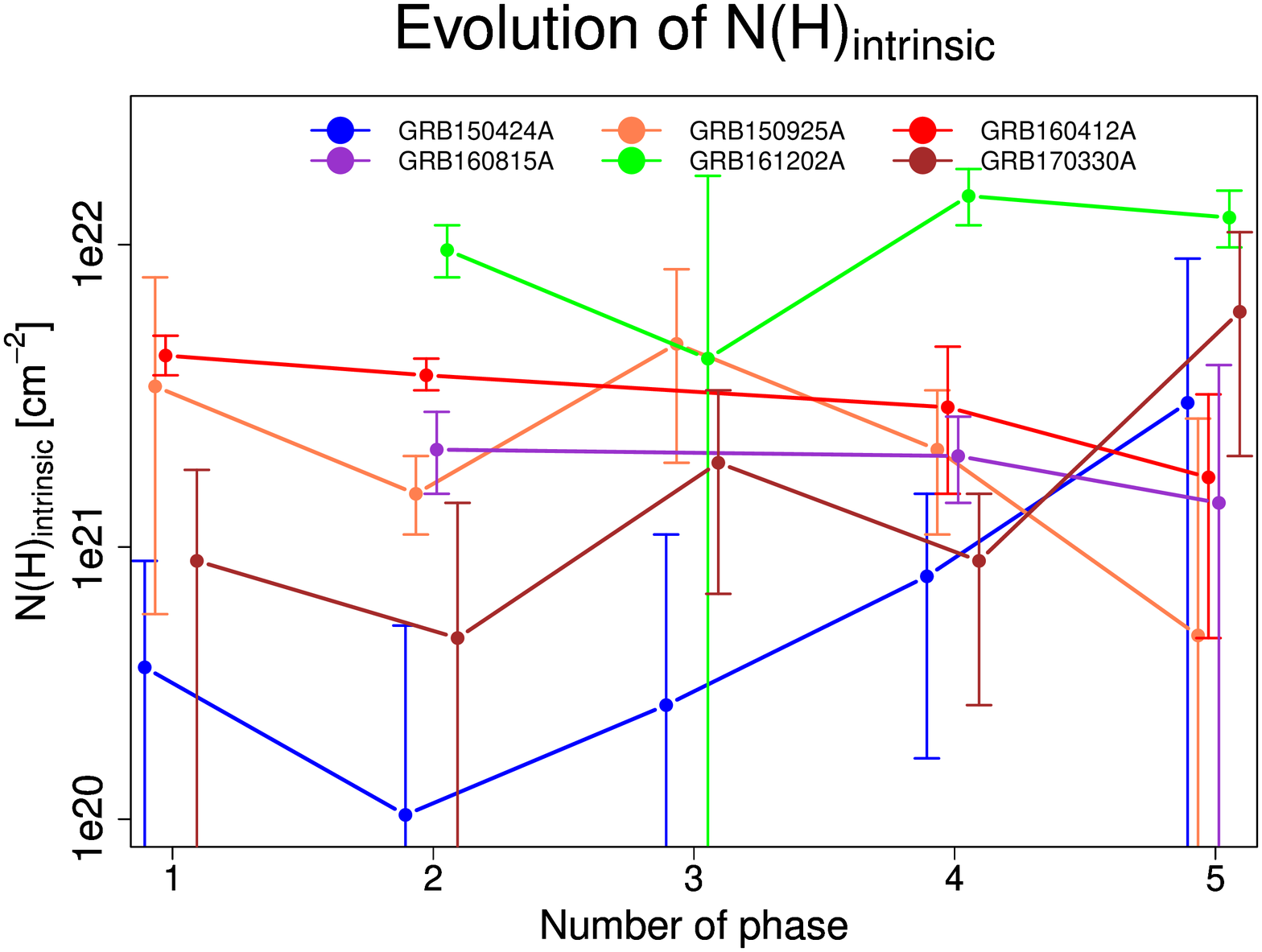}
\caption{The fitted intrinsic N(H) column densities for the 6 GRBs and for each time phase. The 1 $\sigma$ errorbars are also shown. We applied slight horizontal shifts for the different GRBs on the X axis for better visibility. \label{phase}}
\end{figure}

To check this scenario we performed the spectral fitting for the chosen 6 sample GRBs on each 4+1 phase of the spectra (in both WT and PC mode). The intrinsic N(H) is typically between the $10^{21}$ and $10^{23}$ $\textrm{cm}^{-2}$ as was published in Sec.~4.2.~\citet{evans2009} and the 6 chosen GRBs can be considered typical. The results are shown in Fig.~\ref{phase}. In contrast to our theoretical expectations, it does not show any significant decrease in the intrinsic column density (in time). The total evolution of the intrinsic column density can be seen in Table~\ref{evo2}. and a typical deviation of the fitting. Because the errors of fittings have the same order of magnitude as the calculated intrinsic column densities, the variability determined by 1 $\sigma$ statistical errors is too wide. We can therefore declare that it is not a convincing evidence for the column density decrease.

\begin{table}
\centering
\caption{We checked the difference between the first and last NH intrinsic values. Three out of six times the values were found to be decreasing. The typical 1 $\sigma$ errors were comparable to the varying of the N(H) and therefore it cannot be called a significant effect. \label{evo2}}
\begin{tabular}{c|c|c}
ID of GRB&$\Delta$N(H)$_{\textrm{intrinsic}}$&Median errors of \\
& [cm$^{-2}$]& the fitting [cm$^{-2}$]\\ \hline
GRB150424A&$+2.6\cdot 10^{21}$& $0.55 \cdot 10^{21}$\\
GRB150925A&$-2.9\cdot 10^{21}$& $1.33 \cdot 10^{21}$\\
GRB160412A&$-2.6\cdot 10^{21}$& $1.00 \cdot 10^{21}$\\
GRB160815A&$-0.7\cdot 10^{21}$& $0.65 \cdot 10^{21}$\\
GRB161202A&$+2.7\cdot 10^{21}$& $3.05 \cdot 10^{21}$\\
GRB170330A&$+5.1\cdot 10^{21}$& $0.70 \cdot 10^{21}$\\
\end{tabular}
\end{table}

\begin{figure}
\includegraphics[width=0.99\linewidth, trim= 0 0 0 0]{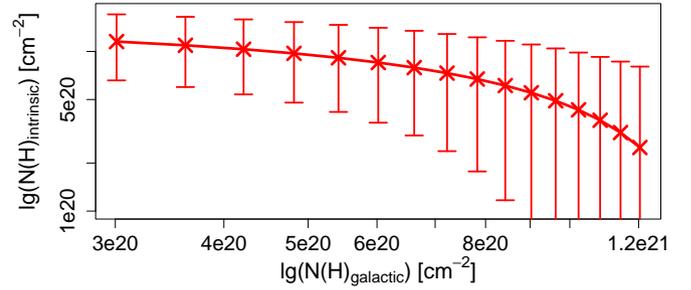}
\caption{An example of the variance of the N(H) intrinsic column density for GRB150424A.  The X-ray Time-averaged spectrum was fitted, while changing the Galactic foreground between 50\% and 200\% of the standard values (LAB Survey). Changing the Galactic foreground yields the same magnitude variance as the evolution showed on Fig.~\ref{phase} where the intrinsic and the foreground column densities are commensurate. \label{robust}}
\end{figure}

We also checked the robustness of the fitting. It strongly depends on the galactic foreground density, therefore we varied the galactic foreground values in the range of 50\% to 200\% of the classic values based on LAB Survey \citep{lab}. Fig.~\ref{robust}. shows an example of the N(H) intrinsic column density variability. 
The variance of the intrinsic N(H) are about $1\cdot 10^{21} \textrm{cm}^{-2}$, which means more than 105 percent error of the original values.
If the fitted intrinsic N(H) is an order of magnitude higher than the Galactic foreground N(H) the variability from the foreground is negligible. 
The time evolution of the intrinsic N(H) shows that a precise galactic foreground estimation is essential, because that might significantly modify the calculated values in certain cases.
We also note that as the galactic foreground seems to show filamentary structures as it was published by \citet{toth1}. This structure can cause over a magnitude of changes in the measured column density in small scales.

Results suggest that the variability of the calculated intrinsic column densities was smaller than the statistical errors of the fitting.






\section{Summary}
\label{sum}



We analyzed the X-ray spectra of the 6 chosen GRBs that have 3 breakpoints in the light curves.  The fitted intrinsic N(H) has linear dependence on the galactic foreground and quadratic relation with the redshift.  The galactic ISM shows fine density structures, therefore the precise (high-resolution) N(H) hydrogen column density is necessary to preform accurate fitting. Our results are consistent with the theoretical expectations. 

The majority of the observed X-ray radiation originates from the interaction between the jet and the interstellar matter. The jet advances in the matter, and as the matter gets thinner, the modification of the created X-ray photons can change. Consequently, the intrinsic column density could vary in time. The errors of  the fitting of the Swift X-ray spectra were too big to show the variability of the intrinsic N(H) column densities.

\section*{Acknowledgments}

Supported by the \fundingAgency{Hungarian OTKA} \fundingNumber{NN-111016} grant and supported by the New National Excellence Program of the \fundingAgency{Hungarian Ministry of Human Capacities} \fundingNumber{UNKP-17-3}. We are grateful to Lajos G. Balazs and Zsolt Bagoly for the enlightening discussions. We are thankful to the Anonymous Referee for his helpful suggestions concerning the presentation of this paper.

\subsection*{Author contributions}
All authors participated in acquisition, analysis and interpretation of data as well as the preparation of the manuscript.

\subsection*{Financial disclosure}

None reported.

\subsection*{Conflict of interest}

The authors declare no potential conflict of interests.

\bibliography{hortobagyi}

\begin{thebibliography}{}

\bibitem [\protect \citeauthoryear {%
{Amati}%
\ \protect \BOthers {.}}{%
{Amati}%
\ \protect \BOthers {.}}{%
{\protect \APACyear {2013}}%
}]{%
amati}
\APACinsertmetastar {%
amati}%
\begin{APACrefauthors}%
{Amati}, L.%
, {Atteia}, J\BHBI L.%
, {Balazs}, L.%
\ et al.\end{APACrefauthors}%
\unskip\
\newblock
\APACrefYearMonthDay{2013}{{\APACmonth{06}}}{},
\newblock
\unskip
\newblock
\APACjournalVolNumPages{White paper}{}{}{arXiv:1306.5259}.
\PrintBackRefs{\CurrentBib}

\bibitem [\protect \citeauthoryear {%
K.~{Arnaud}%
, {Smith}%
\BCBL {}\ \BBA {} {Siemiginowska}%
}{%
K.~{Arnaud}%
\ \protect \BOthers {.}}{%
{\protect \APACyear {2011}}%
}]{%
arnaud}
\APACinsertmetastar {%
arnaud}%
\begin{APACrefauthors}%
{Arnaud}, K.%
, {Smith}, R.%
\BCBL {}\ \BBA {} {Siemiginowska}, A.%
\end{APACrefauthors}%
\unskip\
\newblock
\APACrefYear{2011},
\newblock
\APACrefbtitle {{Handbook of X-ray Astronomy}} {{Handbook of X-ray Astronomy}}.
\PrintBackRefs{\CurrentBib}

\bibitem [\protect \citeauthoryear {%
K\BPBI A.~{Arnaud}%
}{%
K\BPBI A.~{Arnaud}%
}{%
{\protect \APACyear {1996}}%
}]{%
xpsec1}
\APACinsertmetastar {%
xpsec1}%
\begin{APACrefauthors}%
{Arnaud}, K\BPBI A.%
\end{APACrefauthors}%
\unskip\
\newblock
\APACrefYearMonthDay{1996}{}{},
\newblock
{\BBOQ}\APACrefatitle {{XSPEC: The First Ten Years}} {{XSPEC: The First Ten
  Years}}.{\BBCQ}
\newblock
\BIn{} G\BPBI H.~{Jacoby}\ \BBA {} J.~{Barnes}\ (\BEDS), \APACrefbtitle
  {Astron. Data Anal. Soft. and Sys. V} {Astron. Data Anal. Soft. and Sys. V}\
  \BVOL~101, \BPG~17.
\PrintBackRefs{\CurrentBib}

\bibitem [\protect \citeauthoryear {%
{Bagoly}%
\ \protect \BOthers {.}}{%
{Bagoly}%
\ \protect \BOthers {.}}{%
{\protect \APACyear {2003}}%
{\protect \APACexlab {{\protect \BCnt {1}}}}}]{%
zsolt2003_redshift}
\APACinsertmetastar {%
zsolt2003_redshift}%
\begin{APACrefauthors}%
{Bagoly}, Z.%
, {Csabai}, I.%
, {M{\'e}sz{\'a}ros}, A.%
, {M{\'e}sz{\'a}ros}, P.%
, {Horv{\'a}th}, I.%
, {Bal{\'a}zs}, L\BPBI G.%
\BCBL {}\ \BBA {} {Vavrek}, R.%
\end{APACrefauthors}%
\unskip\
\newblock
\APACrefYearMonthDay{2003{\protect \BCnt {1}}}{{\APACmonth{04}}}{},
\newblock
{\BBOQ}\APACrefatitle {{Estimation of the Redshifts for Long Gamma-Ray Bursts}}
  {{Estimation of the Redshifts for Long Gamma-Ray Bursts}}.{\BBCQ}
\newblock
\BIn{} \APACrefbtitle {GAMMA-RAY BURST AND AFTERGLOW ASTRONOMY 2001: A Workshop
  Celebrating the First Year of the HETE Mission. AIP Conference Proceedings,
  Volume 662, pp. 446-449 (2003).} {GAMMA-RAY BURST AND AFTERGLOW ASTRONOMY
  2001: A Workshop Celebrating the First Year of the HETE Mission. AIP
  Conference Proceedings, Volume 662, pp. 446-449 (2003).}\ \BVOL~662,
  \BPG~446-449.
\newblock
\begin{APACrefDOI} \doi{10.1063/1.1579398} \end{APACrefDOI}
\PrintBackRefs{\CurrentBib}

\bibitem [\protect \citeauthoryear {%
{Bagoly}%
\ \protect \BOthers {.}}{%
{Bagoly}%
\ \protect \BOthers {.}}{%
{\protect \APACyear {2003}}%
{\protect \APACexlab {{\protect \BCnt {2}}}}}]{%
zsolt2003_redshift2}
\APACinsertmetastar {%
zsolt2003_redshift2}%
\begin{APACrefauthors}%
{Bagoly}, Z.%
, {Csabai}, I.%
, {M{\'e}sz{\'a}ros}, A.%
, {M{\'e}sz{\'a}ros}, P.%
, {Horv{\'a}th}, I.%
, {Bal{\'a}zs}, L\BPBI G.%
\BCBL {}\ \BBA {} {Vavrek}, R.%
\end{APACrefauthors}%
\unskip\
\newblock
\APACrefYearMonthDay{2003{\protect \BCnt {2}}}{{\APACmonth{02}}}{},
\newblock
\unskip
\newblock
\APACjournalVolNumPages{\aap}{398}{}{919-925}.
\newblock
\begin{APACrefDOI} \doi{10.1051/0004-6361:20021724} \end{APACrefDOI}
\PrintBackRefs{\CurrentBib}

\bibitem [\protect \citeauthoryear {%
{Bagoly}%
\ \protect \BOthers {.}}{%
{Bagoly}%
\ \protect \BOthers {.}}{%
{\protect \APACyear {2006}}%
}]{%
zsolt2006_redshift}
\APACinsertmetastar {%
zsolt2006_redshift}%
\begin{APACrefauthors}%
{Bagoly}, Z.%
, {M{\'e}sz{\'a}ros}, A.%
, {Bal{\'a}zs}, L\BPBI G.%
\ et al.\end{APACrefauthors}%
\unskip\
\newblock
\APACrefYearMonthDay{2006}{{\APACmonth{07}}}{},
\newblock
\unskip
\newblock
\APACjournalVolNumPages{\aap}{453}{}{797-800}.
\newblock
\begin{APACrefDOI} \doi{10.1051/0004-6361:20054322} \end{APACrefDOI}
\PrintBackRefs{\CurrentBib}

\bibitem [\protect \citeauthoryear {%
{Bagoly}%
, {M{\'e}sz{\'a}ros}%
, {Horv{\'a}th}%
, {Bal{\'a}zs}%
\BCBL {}\ \BBA {} {M{\'e}sz{\'a}ros}%
}{%
{Bagoly}%
\ \protect \BOthers {.}}{%
{\protect \APACyear {1998}}%
}]{%
zsolt1998_pca}
\APACinsertmetastar {%
zsolt1998_pca}%
\begin{APACrefauthors}%
{Bagoly}, Z.%
, {M{\'e}sz{\'a}ros}, A.%
, {Horv{\'a}th}, I.%
, {Bal{\'a}zs}, L\BPBI G.%
\BCBL {}\ \BBA {} {M{\'e}sz{\'a}ros}, P.%
\end{APACrefauthors}%
\unskip\
\newblock
\APACrefYearMonthDay{1998}{{\APACmonth{05}}}{},
\newblock
\unskip
\newblock
\APACjournalVolNumPages{\apj}{498}{}{342-348}.
\newblock
\begin{APACrefDOI} \doi{10.1086/305530} \end{APACrefDOI}
\PrintBackRefs{\CurrentBib}

\bibitem [\protect \citeauthoryear {%
{Bal{\'a}zs}%
\ \protect \BOthers {.}}{%
{Bal{\'a}zs}%
\ \protect \BOthers {.}}{%
{\protect \APACyear {2015}}%
}]{%
ring}
\APACinsertmetastar {%
ring}%
\begin{APACrefauthors}%
{Bal{\'a}zs}, L\BPBI G.%
, {Bagoly}, Z.%
, {Hakkila}, J\BPBI E.%
, {Horv{\'a}th}, I.%
, {K{\'o}bori}, J.%
, {R{\'a}cz}, I\BPBI I.%
\BCBL {}\ \BBA {} {T{\'o}th}, L\BPBI V.%
\end{APACrefauthors}%
\unskip\
\newblock
\APACrefYearMonthDay{2015}{{\APACmonth{09}}}{},
\newblock
\unskip
\newblock
\APACjournalVolNumPages{\mnras}{452}{}{2236-2246}.
\newblock
\begin{APACrefDOI} \doi{10.1093/mnras/stv1421} \end{APACrefDOI}
\PrintBackRefs{\CurrentBib}

\bibitem [\protect \citeauthoryear {%
{Bal{\'a}zs}%
, {M{\'e}sz{\'a}ros}%
, {Horv{\'a}th}%
\BCBL {}\ \BBA {} {Vavrek}%
}{%
{Bal{\'a}zs}%
\ \protect \BOthers {.}}{%
{\protect \APACyear {1999}}%
}]{%
balazs1999}
\APACinsertmetastar {%
balazs1999}%
\begin{APACrefauthors}%
{Bal{\'a}zs}, L\BPBI G.%
, {M{\'e}sz{\'a}ros}, A.%
, {Horv{\'a}th}, I.%
\BCBL {}\ \BBA {} {Vavrek}, R.%
\end{APACrefauthors}%
\unskip\
\newblock
\APACrefYearMonthDay{1999}{{\APACmonth{09}}}{},
\newblock
\unskip
\newblock
\APACjournalVolNumPages{\aaps}{138}{}{417-418}.
\newblock
\begin{APACrefDOI} \doi{10.1051/aas:1999290} \end{APACrefDOI}
\PrintBackRefs{\CurrentBib}

\bibitem [\protect \citeauthoryear {%
{Bal{\'a}zs}%
, {Rejt{\'{o}}}%
\BCBL {}\ \BBA {} {Tusn{\'a}dy}%
}{%
{Bal{\'a}zs}%
\ \protect \BOthers {.}}{%
{\protect \APACyear {2018}}%
}]{%
ring2}
\APACinsertmetastar {%
ring2}%
\begin{APACrefauthors}%
{Bal{\'a}zs}, L\BPBI G.%
, {Rejt{\'{o}}}, L.%
\BCBL {}\ \BBA {} {Tusn{\'a}dy}, G.%
\end{APACrefauthors}%
\unskip\
\newblock
\APACrefYearMonthDay{2018}{{\APACmonth{01}}}{},
\newblock
\unskip
\newblock
\APACjournalVolNumPages{\mnras}{473}{}{3169-3179}.
\newblock
\begin{APACrefDOI} \doi{10.1093/mnras/stx2550} \end{APACrefDOI}
\PrintBackRefs{\CurrentBib}

\bibitem [\protect \citeauthoryear {%
{Barthelmy}%
\ \protect \BOthers {.}}{%
{Barthelmy}%
\ \protect \BOthers {.}}{%
{\protect \APACyear {2005}}%
}]{%
bat}
\APACinsertmetastar {%
bat}%
\begin{APACrefauthors}%
{Barthelmy}, S\BPBI D.%
, {Barbier}, L\BPBI M.%
, {Cummings}, J\BPBI R.%
\ et al.\end{APACrefauthors}%
\unskip\
\newblock
\APACrefYearMonthDay{2005}{{\APACmonth{10}}}{},
\newblock
\unskip
\newblock
\APACjournalVolNumPages{\ssr}{120}{}{143-164}.
\newblock
\begin{APACrefDOI} \doi{10.1007/s11214-005-5096-3} \end{APACrefDOI}
\PrintBackRefs{\CurrentBib}

\bibitem [\protect \citeauthoryear {%
{Born}%
}{%
{Born}%
}{%
{\protect \APACyear {1926}}%
}]{%
born}
\APACinsertmetastar {%
born}%
\begin{APACrefauthors}%
{Born}, M.%
\end{APACrefauthors}%
\unskip\
\newblock
\APACrefYearMonthDay{1926}{{\APACmonth{11}}}{},
\newblock
\unskip
\newblock
\APACjournalVolNumPages{Zeitschrift fur Physik}{38}{}{803-827}.
\newblock
\begin{APACrefDOI} \doi{10.1007/BF01397184} \end{APACrefDOI}
\PrintBackRefs{\CurrentBib}

\bibitem [\protect \citeauthoryear {%
{Burrows}%
\ \protect \BOthers {.}}{%
{Burrows}%
\ \protect \BOthers {.}}{%
{\protect \APACyear {2005}}%
}]{%
xrt}
\APACinsertmetastar {%
xrt}%
\begin{APACrefauthors}%
{Burrows}, D\BPBI N.%
, {Hill}, J\BPBI E.%
, {Nousek}, J\BPBI A.%
\ et al.\end{APACrefauthors}%
\unskip\
\newblock
\APACrefYearMonthDay{2005}{{\APACmonth{10}}}{},
\newblock
\unskip
\newblock
\APACjournalVolNumPages{\ssr}{120}{}{165-195}.
\newblock
\begin{APACrefDOI} \doi{10.1007/s11214-005-5097-2} \end{APACrefDOI}
\PrintBackRefs{\CurrentBib}

\bibitem [\protect \citeauthoryear {%
{de Ugarte Postigo}%
\ \protect \BOthers {.}}{%
{de Ugarte Postigo}%
\ \protect \BOthers {.}}{%
{\protect \APACyear {2011}}%
}]{%
ugarte}
\APACinsertmetastar {%
ugarte}%
\begin{APACrefauthors}%
{de Ugarte Postigo}, A.%
, {Horv{\'a}th}, I.%
, {Veres}, P.%
\ et al.\end{APACrefauthors}%
\unskip\
\newblock
\APACrefYearMonthDay{2011}{{\APACmonth{01}}}{},
\newblock
\unskip
\newblock
\APACjournalVolNumPages{\aap}{525}{}{A109}.
\newblock
\begin{APACrefDOI} \doi{10.1051/0004-6361/201015261} \end{APACrefDOI}
\PrintBackRefs{\CurrentBib}

\bibitem [\protect \citeauthoryear {%
{Dorman}%
\ \BBA {} {Arnaud}%
}{%
{Dorman}%
\ \BBA {} {Arnaud}%
}{%
{\protect \APACyear {2001}}%
}]{%
xpsec2}
\APACinsertmetastar {%
xpsec2}%
\begin{APACrefauthors}%
{Dorman}, B.%
\BCBT {}\ \BBA {} {Arnaud}, K\BPBI A.%
\end{APACrefauthors}%
\unskip\
\newblock
\APACrefYearMonthDay{2001}{}{},
\newblock
{\BBOQ}\APACrefatitle {{Redesign and Reimplementation of XSPEC}} {{Redesign and
  Reimplementation of XSPEC}}.{\BBCQ}
\newblock
\BIn{} F\BPBI R.~{Harnden} Jr., F\BPBI A.~{Primini}\BCBL {}\ \BBA {} H\BPBI
  E.~{Payne}\ (\BEDS), \APACrefbtitle {Astron. Data Anal. Soft. and Sys. X}
  {Astron. Data Anal. Soft. and Sys. X}\ \BVOL~238, \BPG~415.
\PrintBackRefs{\CurrentBib}

\bibitem [\protect \citeauthoryear {%
{Dorman}%
, {Arnaud}%
\BCBL {}\ \BBA {} {Gordon}%
}{%
{Dorman}%
\ \protect \BOthers {.}}{%
{\protect \APACyear {2003}}%
}]{%
xpsec3}
\APACinsertmetastar {%
xpsec3}%
\begin{APACrefauthors}%
{Dorman}, B.%
, {Arnaud}, K\BPBI A.%
\BCBL {}\ \BBA {} {Gordon}, C\BPBI A.%
\end{APACrefauthors}%
\unskip\
\newblock
\APACrefYearMonthDay{2003}{{\APACmonth{03}}}{},
\newblock
{\BBOQ}\APACrefatitle {{XSPEC12: Object-Oriented X-Ray Analysis}} {{XSPEC12:
  Object-Oriented X-Ray Analysis}}.{\BBCQ}
\newblock
\BIn{} \APACrefbtitle {AAS/High Energy Astrophysics Division \#7} {AAS/High
  Energy Astrophysics Division \#7}\ \BVOL~35, \BPG~641.
\PrintBackRefs{\CurrentBib}

\bibitem [\protect \citeauthoryear {%
{Evans}%
\ \protect \BOthers {.}}{%
{Evans}%
\ \protect \BOthers {.}}{%
{\protect \APACyear {2007}}%
}]{%
evans2007}
\APACinsertmetastar {%
evans2007}%
\begin{APACrefauthors}%
{Evans}, P\BPBI A.%
, { Beardmore}, A\BPBI P.%
, { Page}, K\BPBI L.%
\ et al.\end{APACrefauthors}%
\unskip\
\newblock
\APACrefYearMonthDay{2007}{{\APACmonth{07}}}{},
\newblock
\unskip
\newblock
\APACjournalVolNumPages{\aaa}{469}{}{379–385}.
\newblock
\begin{APACrefDOI} \doi{10.1051/0004-6361:20077530} \end{APACrefDOI}
\PrintBackRefs{\CurrentBib}

\bibitem [\protect \citeauthoryear {%
{Evans}%
\ \protect \BOthers {.}}{%
{Evans}%
\ \protect \BOthers {.}}{%
{\protect \APACyear {2009}}%
}]{%
evans2009}
\APACinsertmetastar {%
evans2009}%
\begin{APACrefauthors}%
{Evans}, P\BPBI A.%
, {Beardmore}, A\BPBI P.%
, {Page}, K\BPBI L.%
\ et al.\end{APACrefauthors}%
\unskip\
\newblock
\APACrefYearMonthDay{2009}{{\APACmonth{07}}}{},
\newblock
\unskip
\newblock
\APACjournalVolNumPages{\mnras}{397}{}{1177-1201}.
\newblock
\begin{APACrefDOI} \doi{10.1111/j.1365-2966.2009.14913.x} \end{APACrefDOI}
\PrintBackRefs{\CurrentBib}

\bibitem [\protect \citeauthoryear {%
{Horv{\'a}th}%
}{%
{Horv{\'a}th}%
}{%
{\protect \APACyear {1998}}%
}]{%
hoi1998}
\APACinsertmetastar {%
hoi1998}%
\begin{APACrefauthors}%
{Horv{\'a}th}, I.%
\end{APACrefauthors}%
\unskip\
\newblock
\APACrefYearMonthDay{1998}{{\APACmonth{12}}}{},
\newblock
\unskip
\newblock
\APACjournalVolNumPages{\apj}{508}{}{757-759}.
\newblock
\begin{APACrefDOI} \doi{10.1086/306416} \end{APACrefDOI}
\PrintBackRefs{\CurrentBib}

\bibitem [\protect \citeauthoryear {%
{Horv{\'a}th}%
}{%
{Horv{\'a}th}%
}{%
{\protect \APACyear {2002}}%
}]{%
hoi2002}
\APACinsertmetastar {%
hoi2002}%
\begin{APACrefauthors}%
{Horv{\'a}th}, I.%
\end{APACrefauthors}%
\unskip\
\newblock
\APACrefYearMonthDay{2002}{{\APACmonth{09}}}{},
\newblock
\unskip
\newblock
\APACjournalVolNumPages{\aap}{392}{}{791-793}.
\newblock
\begin{APACrefDOI} \doi{10.1051/0004-6361:20020808} \end{APACrefDOI}
\PrintBackRefs{\CurrentBib}

\bibitem [\protect \citeauthoryear {%
{Horv{\'a}th}%
\ \protect \BOthers {.}}{%
{Horv{\'a}th}%
\ \protect \BOthers {.}}{%
{\protect \APACyear {2010}}%
}]{%
hoi2010}
\APACinsertmetastar {%
hoi2010}%
\begin{APACrefauthors}%
{Horv{\'a}th}, I.%
, {Bagoly}, Z.%
, {Bal{\'a}zs}, L\BPBI G.%
, {de Ugarte Postigo}, A.%
, {Veres}, P.%
\BCBL {}\ \BBA {} {M{\'e}sz{\'a}ros}, A.%
\end{APACrefauthors}%
\unskip\
\newblock
\APACrefYearMonthDay{2010}{{\APACmonth{04}}}{},
\newblock
\unskip
\newblock
\APACjournalVolNumPages{\apj}{713}{}{552-557}.
\newblock
\begin{APACrefDOI} \doi{10.1088/0004-637X/713/1/552} \end{APACrefDOI}
\PrintBackRefs{\CurrentBib}

\bibitem [\protect \citeauthoryear {%
{Horv{\'a}th}%
, {Bagoly}%
, {Hakkila}%
\BCBL {}\ \BBA {} {T{\'o}th}%
}{%
{Horv{\'a}th}%
\ \protect \BOthers {.}}{%
{\protect \APACyear {2015}}%
}]{%
hoi2015}
\APACinsertmetastar {%
hoi2015}%
\begin{APACrefauthors}%
{Horv{\'a}th}, I.%
, {Bagoly}, Z.%
, {Hakkila}, J.%
\BCBL {}\ \BBA {} {T{\'o}th}, L\BPBI V.%
\end{APACrefauthors}%
\unskip\
\newblock
\APACrefYearMonthDay{2015}{{\APACmonth{12}}}{},
\newblock
\unskip
\newblock
\APACjournalVolNumPages{\aaa}{584}{}{A48}.
\newblock
\begin{APACrefDOI} \doi{10.1051/0004-6361/201424829} \end{APACrefDOI}
\PrintBackRefs{\CurrentBib}

\bibitem [\protect \citeauthoryear {%
{Horv{\'a}th}%
, {Bal{\'a}zs}%
, {Bagoly}%
, {Ryde}%
\BCBL {}\ \BBA {} {M{\'e}sz{\'a}ros}%
}{%
{Horv{\'a}th}%
\ \protect \BOthers {.}}{%
{\protect \APACyear {2006}}%
}]{%
hoi2006}
\APACinsertmetastar {%
hoi2006}%
\begin{APACrefauthors}%
{Horv{\'a}th}, I.%
, {Bal{\'a}zs}, L\BPBI G.%
, {Bagoly}, Z.%
, {Ryde}, F.%
\BCBL {}\ \BBA {} {M{\'e}sz{\'a}ros}, A.%
\end{APACrefauthors}%
\unskip\
\newblock
\APACrefYearMonthDay{2006}{{\APACmonth{02}}}{},
\newblock
\unskip
\newblock
\APACjournalVolNumPages{\aap}{447}{}{23-30}.
\newblock
\begin{APACrefDOI} \doi{10.1051/0004-6361:20041129} \end{APACrefDOI}
\PrintBackRefs{\CurrentBib}

\bibitem [\protect \citeauthoryear {%
{Horv{\'a}th}%
, {Bal{\'a}zs}%
, {Bagoly}%
\BCBL {}\ \BBA {} {Veres}%
}{%
{Horv{\'a}th}%
\ \protect \BOthers {.}}{%
{\protect \APACyear {2008}}%
}]{%
hoi2008}
\APACinsertmetastar {%
hoi2008}%
\begin{APACrefauthors}%
{Horv{\'a}th}, I.%
, {Bal{\'a}zs}, L\BPBI G.%
, {Bagoly}, Z.%
\BCBL {}\ \BBA {} {Veres}, P.%
\end{APACrefauthors}%
\unskip\
\newblock
\APACrefYearMonthDay{2008}{{\APACmonth{10}}}{},
\newblock
\unskip
\newblock
\APACjournalVolNumPages{\aaa}{489}{}{L1-L4}.
\newblock
\begin{APACrefDOI} \doi{10.1051/0004-6361:200810269} \end{APACrefDOI}
\PrintBackRefs{\CurrentBib}

\bibitem [\protect \citeauthoryear {%
{Horv{\'a}th}%
, {Hakkila}%
\BCBL {}\ \BBA {} {Bagoly}%
}{%
{Horv{\'a}th}%
\ \protect \BOthers {.}}{%
{\protect \APACyear {2014}}%
}]{%
hoi2014}
\APACinsertmetastar {%
hoi2014}%
\begin{APACrefauthors}%
{Horv{\'a}th}, I.%
, {Hakkila}, J.%
\BCBL {}\ \BBA {} {Bagoly}, Z.%
\end{APACrefauthors}%
\unskip\
\newblock
\APACrefYearMonthDay{2014}{{\APACmonth{01}}}{},
\newblock
\unskip
\newblock
\APACjournalVolNumPages{\aaa}{561}{}{L12}.
\newblock
\begin{APACrefDOI} \doi{10.1051/0004-6361/201323020} \end{APACrefDOI}
\PrintBackRefs{\CurrentBib}

\bibitem [\protect \citeauthoryear {%
{Horv{\'a}th}%
\ \protect \BOthers {.}}{%
{Horv{\'a}th}%
\ \protect \BOthers {.}}{%
{\protect \APACyear {2018}}%
}]{%
hoi2018}
\APACinsertmetastar {%
hoi2018}%
\begin{APACrefauthors}%
{Horv{\'a}th}, I.%
, {T{\'o}th}, B\BPBI G.%
, {Hakkila}, J.%
\ et al.\end{APACrefauthors}%
\unskip\
\newblock
\APACrefYearMonthDay{2018}{{\APACmonth{03}}}{},
\newblock
\unskip
\newblock
\APACjournalVolNumPages{\apss}{363}{}{}.
\newblock
\begin{APACrefDOI} \doi{10.1007/s10509-018-3274-5} \end{APACrefDOI}
\PrintBackRefs{\CurrentBib}

\bibitem [\protect \citeauthoryear {%
{Kalberla}%
\ \protect \BOthers {.}}{%
{Kalberla}%
\ \protect \BOthers {.}}{%
{\protect \APACyear {2005}}%
}]{%
lab}
\APACinsertmetastar {%
lab}%
\begin{APACrefauthors}%
{Kalberla}, P\BPBI M\BPBI W.%
, {Burton}, W\BPBI B.%
, {Hartmann}, D.%
, {Arnal}, E\BPBI M.%
, {Bajaja}, E.%
, {Morras}, R.%
\BCBL {}\ \BBA {} {P{\"o}ppel}, W\BPBI G\BPBI L.%
\end{APACrefauthors}%
\unskip\
\newblock
\APACrefYearMonthDay{2005}{{\APACmonth{09}}}{},
\newblock
\unskip
\newblock
\APACjournalVolNumPages{\aaa}{440}{}{775-782}.
\newblock
\begin{APACrefDOI} \doi{10.1051/0004-6361:20041864} \end{APACrefDOI}
\PrintBackRefs{\CurrentBib}

\bibitem [\protect \citeauthoryear {%
{Kumar}%
\ \BBA {} {Zhang}%
}{%
{Kumar}%
\ \BBA {} {Zhang}%
}{%
{\protect \APACyear {2015}}%
}]{%
kumar}
\APACinsertmetastar {%
kumar}%
\begin{APACrefauthors}%
{Kumar}, P.%
\BCBT {}\ \BBA {} {Zhang}, B.%
\end{APACrefauthors}%
\unskip\
\newblock
\APACrefYearMonthDay{2015}{{\APACmonth{02}}}{},
\newblock
\unskip
\newblock
\APACjournalVolNumPages{Physics Reports}{561}{}{1-109}.
\newblock
\begin{APACrefDOI} \doi{10.1016/j.physrep.2014.09.008} \end{APACrefDOI}
\PrintBackRefs{\CurrentBib}

\bibitem [\protect \citeauthoryear {%
A.~{M{\'e}sz{\'a}ros}%
, {Bagoly}%
, {Horv{\'a}th}%
, {Bal{\'a}zs}%
\BCBL {}\ \BBA {} {Vavrek}%
}{%
A.~{M{\'e}sz{\'a}ros}%
, {Bagoly}%
, {Horv{\'a}th}%
\BCBL {}\ \protect \BOthers {.}}{%
{\protect \APACyear {2000}}%
}]{%
zsolt2000_anizotrop}
\APACinsertmetastar {%
zsolt2000_anizotrop}%
\begin{APACrefauthors}%
{M{\'e}sz{\'a}ros}, A.%
, {Bagoly}, Z.%
, {Horv{\'a}th}, I.%
, {Bal{\'a}zs}, L\BPBI G.%
\BCBL {}\ \BBA {} {Vavrek}, R.%
\end{APACrefauthors}%
\unskip\
\newblock
\APACrefYearMonthDay{2000}{{\APACmonth{08}}}{},
\newblock
\unskip
\newblock
\APACjournalVolNumPages{\apj}{539}{}{98-101}.
\newblock
\begin{APACrefDOI} \doi{10.1086/309193} \end{APACrefDOI}
\PrintBackRefs{\CurrentBib}

\bibitem [\protect \citeauthoryear {%
A.~{M{\'e}sz{\'a}ros}%
, {Bagoly}%
\BCBL {}\ \BBA {} {Vavrek}%
}{%
A.~{M{\'e}sz{\'a}ros}%
, {Bagoly}%
\BCBL {}\ \BBA {} {Vavrek}%
}{%
{\protect \APACyear {2000}}%
}]{%
zsolt2000_anizotrop2}
\APACinsertmetastar {%
zsolt2000_anizotrop2}%
\begin{APACrefauthors}%
{M{\'e}sz{\'a}ros}, A.%
, {Bagoly}, Z.%
\BCBL {}\ \BBA {} {Vavrek}, R.%
\end{APACrefauthors}%
\unskip\
\newblock
\APACrefYearMonthDay{2000}{{\APACmonth{02}}}{},
\newblock
\unskip
\newblock
\APACjournalVolNumPages{\aap}{354}{}{1-6}.
\PrintBackRefs{\CurrentBib}

\bibitem [\protect \citeauthoryear {%
P.~{M{\'e}sz{\'a}ros}%
}{%
P.~{M{\'e}sz{\'a}ros}%
}{%
{\protect \APACyear {2006}}%
}]{%
meszaros_2006}
\APACinsertmetastar {%
meszaros_2006}%
\begin{APACrefauthors}%
{M{\'e}sz{\'a}ros}, P.%
\end{APACrefauthors}%
\unskip\
\newblock
\APACrefYearMonthDay{2006}{{\APACmonth{08}}}{},
\newblock
\unskip
\newblock
\APACjournalVolNumPages{Reports on Progress in Physics}{69}{}{2259-2321}.
\newblock
\begin{APACrefDOI} \doi{10.1088/0034-4885/69/8/R01} \end{APACrefDOI}
\PrintBackRefs{\CurrentBib}

\bibitem [\protect \citeauthoryear {%
P.~{M{\'e}sz{\'a}ros}%
, {Asano}%
\BCBL {}\ \BBA {} {Veres}%
}{%
P.~{M{\'e}sz{\'a}ros}%
\ \protect \BOthers {.}}{%
{\protect \APACyear {2014}}%
}]{%
meszaros}
\APACinsertmetastar {%
meszaros}%
\begin{APACrefauthors}%
{M{\'e}sz{\'a}ros}, P.%
, {Asano}, K.%
\BCBL {}\ \BBA {} {Veres}, P.%
\end{APACrefauthors}%
\unskip\
\newblock
\APACrefYearMonthDay{2014}{{\APACmonth{03}}}{},
\newblock
{\BBOQ}\APACrefatitle {{Gamma-ray bursts: Recent results and connections to
  very high energy cosmic rays and neutrinos}} {{Gamma-ray bursts: Recent
  results and connections to very high energy cosmic rays and
  neutrinos}}.{\BBCQ}
\newblock
\BIn{} \APACrefbtitle {Journal of Physics Conference Series} {Journal of
  Physics Conference Series}\ \BVOL~485, \BPG~012001.
\newblock
\begin{APACrefDOI} \doi{10.1088/1742-6596/485/1/012001} \end{APACrefDOI}
\PrintBackRefs{\CurrentBib}

\bibitem [\protect \citeauthoryear {%
{R{\'a}cz}%
\ \protect \BOthers {.}}{%
{R{\'a}cz}%
\ \protect \BOthers {.}}{%
{\protect \APACyear {2017}}%
}]{%
racz}
\APACinsertmetastar {%
racz}%
\begin{APACrefauthors}%
{R{\'a}cz}, I\BPBI I.%
, {Bagoly}, Z.%
, {T{\'o}th}, L\BPBI V.%
, {Bal{\'a}zs}, L\BPBI G.%
, {Horv{\'a}th}, I.%
\BCBL {}\ \BBA {} {Pint{\'e}r}, S.%
\end{APACrefauthors}%
\unskip\
\newblock
\APACrefYearMonthDay{2017}{{\APACmonth{07}}}{},
\newblock
\unskip
\newblock
\APACjournalVolNumPages{Contributions of the Astronomical Observatory Skalnate
  Pleso}{47}{}{100-107}.
\PrintBackRefs{\CurrentBib}

\bibitem [\protect \citeauthoryear {%
{Roming}%
\ \protect \BOthers {.}}{%
{Roming}%
\ \protect \BOthers {.}}{%
{\protect \APACyear {2005}}%
}]{%
uvot}
\APACinsertmetastar {%
uvot}%
\begin{APACrefauthors}%
{Roming}, P\BPBI W\BPBI A.%
, {Kennedy}, T\BPBI E.%
, {Mason}, K\BPBI O.%
\ et al.\end{APACrefauthors}%
\unskip\
\newblock
\APACrefYearMonthDay{2005}{{\APACmonth{10}}}{},
\newblock
\unskip
\newblock
\APACjournalVolNumPages{\ssr}{120}{}{95-142}.
\newblock
\begin{APACrefDOI} \doi{10.1007/s11214-005-5095-4} \end{APACrefDOI}
\PrintBackRefs{\CurrentBib}

\bibitem [\protect \citeauthoryear {%
{Starling}%
\ \protect \BOthers {.}}{%
{Starling}%
\ \protect \BOthers {.}}{%
{\protect \APACyear {2013}}%
}]{%
starling}
\APACinsertmetastar {%
starling}%
\begin{APACrefauthors}%
{Starling}, R\BPBI L\BPBI C.%
, {Willingale}, R.%
, {Tanvir}, N\BPBI R.%
\ et al.\end{APACrefauthors}%
\unskip\
\newblock
\APACrefYearMonthDay{2013}{{\APACmonth{04}}}{},
\newblock
\unskip
\newblock
\APACjournalVolNumPages{\mnras}{431}{}{3159–3176}.
\newblock
\begin{APACrefDOI} \doi{10.1093/mnras/stt400} \end{APACrefDOI}
\PrintBackRefs{\CurrentBib}

\bibitem [\protect \citeauthoryear {%
{Toth}%
\ \protect \BOthers {.}}{%
{Toth}%
\ \protect \BOthers {.}}{%
{\protect \APACyear {2017}}%
}]{%
toth1}
\APACinsertmetastar {%
toth1}%
\begin{APACrefauthors}%
{Toth}, L\BPBI V.%
, {Doi}, Y.%
, {Zahorecz}, S.%
, {Agas}, M.%
, {Balazs}, L\BPBI G.%
, {Forro}, A.%
\BCBL {}\ \BBA {} {Racz}, I\BPBI I.%
\end{APACrefauthors}%
\unskip\
\newblock
\APACrefYearMonthDay{2017}{{\APACmonth{03}}}{},
\newblock
\unskip
\newblock
\APACjournalVolNumPages{Publication of Korean Astronomical
  Society}{32}{}{113-116}.
\newblock
\begin{APACrefDOI} \doi{10.5303/PKAS.2017.32.1.113} \end{APACrefDOI}
\PrintBackRefs{\CurrentBib}

\bibitem [\protect \citeauthoryear {%
{Vavrek}%
, {Bal{\'a}zs}%
, {M{\'e}sz{\'a}ros}%
, {Horv{\'a}th}%
\BCBL {}\ \BBA {} {Bagoly}%
}{%
{Vavrek}%
\ \protect \BOthers {.}}{%
{\protect \APACyear {2008}}%
}]{%
pista_3_anizotrop}
\APACinsertmetastar {%
pista_3_anizotrop}%
\begin{APACrefauthors}%
{Vavrek}, R.%
, {Bal{\'a}zs}, L\BPBI G.%
, {M{\'e}sz{\'a}ros}, A.%
, {Horv{\'a}th}, I.%
\BCBL {}\ \BBA {} {Bagoly}, Z.%
\end{APACrefauthors}%
\unskip\
\newblock
\APACrefYearMonthDay{2008}{{\APACmonth{12}}}{},
\newblock
\unskip
\newblock
\APACjournalVolNumPages{\mnras}{391}{}{1741-1748}.
\newblock
\begin{APACrefDOI} \doi{10.1111/j.1365-2966.2008.13635.x} \end{APACrefDOI}
\PrintBackRefs{\CurrentBib}

\bibitem [\protect \citeauthoryear {%
{Veres}%
, {Bagoly}%
, {Horv{\'a}th}%
, {Bal{\'a}zs}%
\BCBL {}\ \protect \BOthers {.}}{%
{Veres}%
, {Bagoly}%
, {Horv{\'a}th}%
, {Bal{\'a}zs}%
\BCBL {}\ \protect \BOthers {.}}{%
{\protect \APACyear {2010}}%
}]{%
zsolt2010_anizotrop}
\APACinsertmetastar {%
zsolt2010_anizotrop}%
\begin{APACrefauthors}%
{Veres}, P.%
, {Bagoly}, Z.%
, {Horv{\'a}th}, I.%
, {Bal{\'a}zs}, L\BPBI G.%
, {M{\'e}sz{\'a}ros}, A.%
\BCBL {}\ \BBA {} {Kelemen}, J.%
\end{APACrefauthors}%
\unskip\
\newblock
\APACrefYearMonthDay{2010}{{\APACmonth{10}}}{},
\newblock
{\BBOQ}\APACrefatitle {{Directional Anisotropy of Swift Gamma-Ray Bursts}}
  {{Directional Anisotropy of Swift Gamma-Ray Bursts}}.{\BBCQ}
\newblock
\BIn{} \APACrefbtitle {DECIPHERING THE ANCIENT UNIVERSE WITH GAMMA-RAY BURSTS.
  AIP Conference Proceedings, Volume 1279, pp. 457-459 (2010).} {DECIPHERING
  THE ANCIENT UNIVERSE WITH GAMMA-RAY BURSTS. AIP Conference Proceedings,
  Volume 1279, pp. 457-459 (2010).}\ \BVOL\ 1279, \BPG~457-459.
\newblock
\begin{APACrefDOI} \doi{10.1063/1.3509346} \end{APACrefDOI}
\PrintBackRefs{\CurrentBib}

\bibitem [\protect \citeauthoryear {%
{Veres}%
, {Bagoly}%
, {Horv{\'a}th}%
, {M{\'e}sz{\'a}ros}%
\BCBL {}\ \BBA {} {Bal{\'a}zs}%
}{%
{Veres}%
, {Bagoly}%
, {Horv{\'a}th}%
, {M{\'e}sz{\'a}ros}%
\BCBL {}\ \BBA {} {Bal{\'a}zs}%
}{%
{\protect \APACyear {2010}}%
}]{%
pista_2_inter}
\APACinsertmetastar {%
pista_2_inter}%
\begin{APACrefauthors}%
{Veres}, P.%
, {Bagoly}, Z.%
, {Horv{\'a}th}, I.%
, {M{\'e}sz{\'a}ros}, A.%
\BCBL {}\ \BBA {} {Bal{\'a}zs}, L\BPBI G.%
\end{APACrefauthors}%
\unskip\
\newblock
\APACrefYearMonthDay{2010}{{\APACmonth{12}}}{},
\newblock
\unskip
\newblock
\APACjournalVolNumPages{\apj}{725}{}{1955-1964}.
\newblock
\begin{APACrefDOI} \doi{10.1088/0004-637X/725/2/1955} \end{APACrefDOI}
\PrintBackRefs{\CurrentBib}

\bibitem [\protect \citeauthoryear {%
{Verner}%
, {Ferland}%
, {Korista}%
\BCBL {}\ \BBA {} {Yakovlev}%
}{%
{Verner}%
\ \protect \BOthers {.}}{%
{\protect \APACyear {1996}}%
}]{%
vern}
\APACinsertmetastar {%
vern}%
\begin{APACrefauthors}%
{Verner}, D\BPBI A.%
, {Ferland}, G\BPBI J.%
, {Korista}, K\BPBI T.%
\BCBL {}\ \BBA {} {Yakovlev}, D\BPBI G.%
\end{APACrefauthors}%
\unskip\
\newblock
\APACrefYearMonthDay{1996}{{\APACmonth{07}}}{},
\newblock
\unskip
\newblock
\APACjournalVolNumPages{\apj}{465}{}{487}.
\newblock
\begin{APACrefDOI} \doi{10.1086/177435} \end{APACrefDOI}
\PrintBackRefs{\CurrentBib}

\bibitem [\protect \citeauthoryear {%
{Wilms}%
, {Allen}%
\BCBL {}\ \BBA {} {McCray}%
}{%
{Wilms}%
\ \protect \BOthers {.}}{%
{\protect \APACyear {2000}}%
}]{%
wilms}
\APACinsertmetastar {%
wilms}%
\begin{APACrefauthors}%
{Wilms}, J.%
, {Allen}, A.%
\BCBL {}\ \BBA {} {McCray}, R.%
\end{APACrefauthors}%
\unskip\
\newblock
\APACrefYearMonthDay{2000}{{\APACmonth{10}}}{},
\newblock
\unskip
\newblock
\APACjournalVolNumPages{\apj}{542}{}{914-924}.
\newblock
\begin{APACrefDOI} \doi{10.1086/317016} \end{APACrefDOI}
\PrintBackRefs{\CurrentBib}

\bibitem [\protect \citeauthoryear {%
{Zhang}%
, {Yang}%
, {Choi}%
\BCBL {}\ \BBA {} {Chang}%
}{%
{Zhang}%
\ \protect \BOthers {.}}{%
{\protect \APACyear {2016}}%
}]{%
pista1_inter}
\APACinsertmetastar {%
pista1_inter}%
\begin{APACrefauthors}%
{Zhang}, Z\BHBI B.%
, {Yang}, E\BHBI B.%
, {Choi}, C\BHBI S.%
\BCBL {}\ \BBA {} {Chang}, H\BHBI Y.%
\end{APACrefauthors}%
\unskip\
\newblock
\APACrefYearMonthDay{2016}{{\APACmonth{11}}}{},
\newblock
\unskip
\newblock
\APACjournalVolNumPages{\mnras}{462}{}{3243-3254}.
\newblock
\begin{APACrefDOI} \doi{10.1093/mnras/stw1835} \end{APACrefDOI}
\PrintBackRefs{\CurrentBib}

\end{thebibliography}

\section*{Author Biography}
\begin{biography}{\includegraphics[width=60pt,height=70pt]{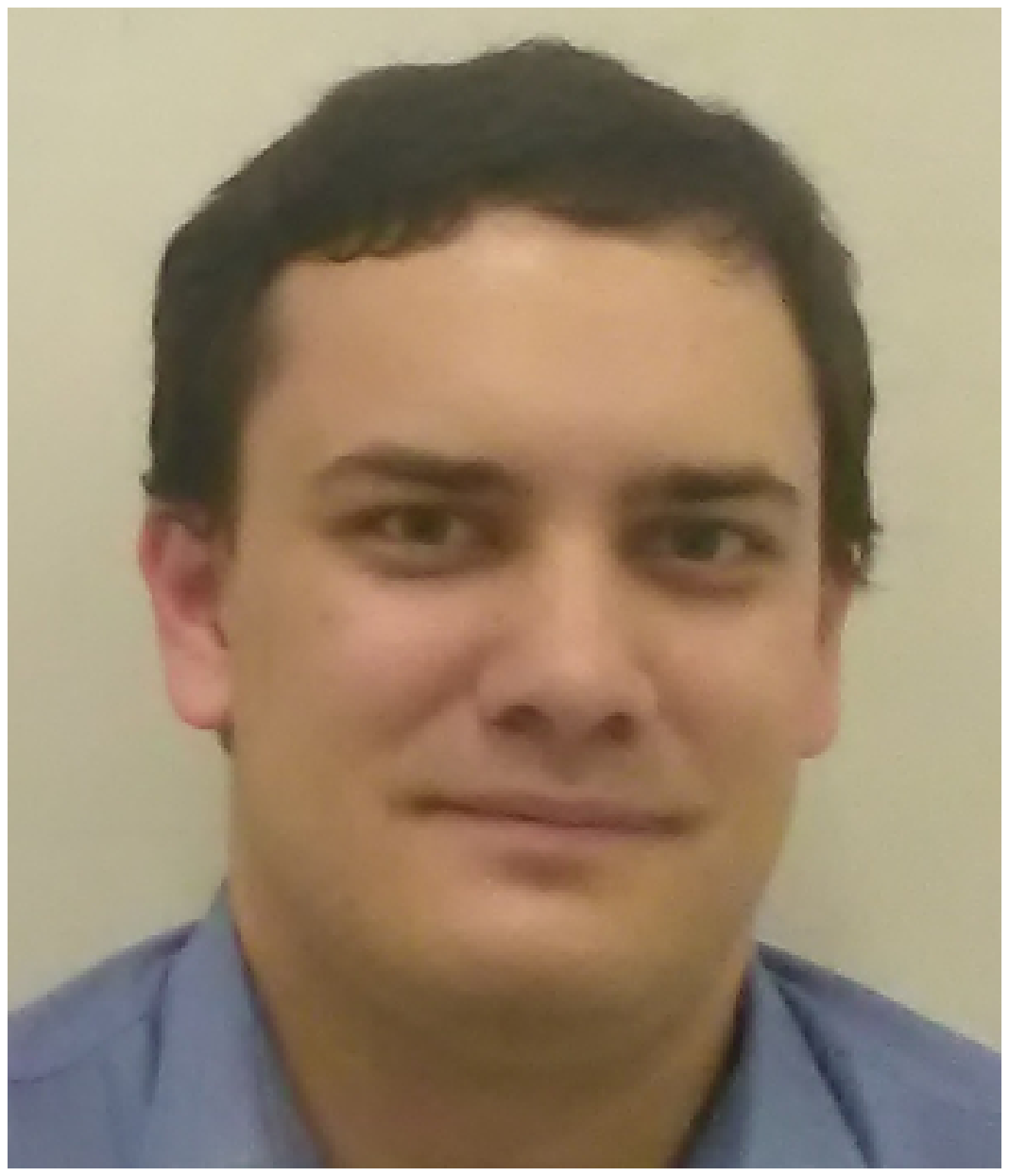}}
{\textbf{Istvan I. Racz.} 
PhD Student, E\"otv\"os Lor\'and University, Budapest, Hungary; Assistant lecturer, National Univesity of Public Services, Budapest, Hungary
Education: PhD School of Physics (Particle Physics and Astronomy), 2015 --, E\"otv\"os Lor\'and University, project: Gamma ray bursts and their Cosmic environment.
MSc in Astronomy, 2015, E\"otv\"os Lor\'and University, thesis: Starformation in Planck cold clumps, supervisor: L. Viktor T\'oth. BSc in Physics, 2012, E\"otv\"os Lor\'and University, thesis: Expansion of the Universe, supervisor: Istvan Csabai.

Research interests: Statistical analysis of the spatial distribution of Gamma-ray bursts, studying the central engine and GRB afterglow with X-ray and Gamma spectral fitting.

}
\end{biography}

\end{document}